\documentclass[12pt]{iopart}

\usepackage{graphicx}
\usepackage{color}
\usepackage[bookmarksnumbered,plainpages]{hyperref}
\begin{document}

\title[Open Problems in Understanding the Nuclear Chirality]{Open Problems in Understanding the Nuclear Chirality}

\author{Jie Meng$^{1,2,3}$, S.Q. Zhang$^{2}$}

 \address{
 $^{1}$ School of Physics and Nuclear Energy Engineering, Beihang University, Beijing 100191, China \\
 $^{2}$ School of Physics and State Key Laboratory of Nuclear Physics
        and Technology, Peking University, 100871 Beijing, China \\
 $^{3}$ Department of Physics, University of Stellenbosch, Stellenbosch, South Africa}
 \ead{mengj@pku.edu.cn}

\begin{abstract}
Open problems in the interpretation of the observed pair of near
degenerate $\Delta I = 1$ bands with the same parity as the chiral
doublet bands are discussed. The ambiguities for the existing
fingerprints of the chirality in atomic nuclei and problems in
existing theory are discussed, including the description of quantum
tunneling in the mean field approximation as well as the
deformation, core polarization and configuration of particle rotor
model (PRM). Future developments of the theoretical approach are
prospected.
\end{abstract}

\pacs{21.60.Ev, 21.60.Jz, 21.10.Re, 23.20.Lv}
\vspace{2pc} \noindent{\it Keywords}: Particle rotor model, Tilted
axis cranking, nuclear chirality,  microscopic theory

\submitto{``Focus issue on Open Problems in Nuclear Structure",
Journal of Physics G}

\maketitle



Handedness or chirality is a subject of general interests in
molecular physics, elementary particles, and optical physics. The
occurrence of chirality in nuclear physics was originally suggested
in 1997 by Frauendorf and Meng in particle-rotor model (PRM) and
tilted axis cranking (TAC) approach for triaxially deformed
nuclei~\cite{FM97}. Since then, lots of
experimental~\cite{{Starosta01,Koike01,Hecht01,Hartley01,Bark01,
LiXF02,Starosta02,Koike03,Rainovski03,Simons05,WangSY06a,
Vaman04,Joshi04,Joshi05,Joshi07,Balabanski04,Lawrie09,Zhu03,
Alcantara04,Timar04,Timar06,Zhao09,Mergel02,ZhuSJ05,LuoYX09,
Tonev06,Petrache06,Grodner06,Mukhopadhyay07,Suzuki08,WangLL09}} as
well as theoretical
efforts~\cite{Frauendorf01,Dimitrov00PRL,Dimitrov00PRC,Olbratowski04,Olbratowski06,
Almehed08,PengJ03,PengJ03cpl,Koike04,ZhangSQ07,WangSY07,WangSY08,
Qi09PLB,Qi09PRC,Higashiyama05,Tonev07,Brant08,Brant09,Droste09,
MengJ06,PengJ08,YaoJM09} have been devoted to search for nuclear
chirality. With the near degenerate doublet bands reported in
$_{55}$Cs, $_{57}$La, $_{59}$Pr and $_{61}$Pm $N=75$ isotones, an
island of chiral rotation was suggested in the $A \sim 130$ mass
region~\cite{Starosta01}. Up to now, candidate chiral doublet bands
have been proposed in a number of odd-odd, odd-$A$ or even-even
nuclei in the $A \sim 100, 130, 190$ mass regions, for a review, see
e.g. Ref.~\cite{MengJ10}.

The best known examples for chiral doublet bands include
$^{126,128}$Cs~\cite{WangSY06a,Grodner06}, $^{106}$Rh~\cite{Joshi04}
and $^{135}$Nd~\cite{Zhu03}, which can be interpreted excellently
with the PRM~\cite{WangSY07,WangSY08,Qi09PLB}. However, a paradox
exists in interpreting a few doublet bands in terms of nuclear
chirality such as $^{134}$Pr~\cite{Tonev06,Petrache06} and
$^{106}$Ag~\cite{Joshi07}, which stimulates the present
investigation.

\section{Nuclear Chirality}

\begin{figure}[h]
\begin{center}
\includegraphics[width=10cm, bb=0 0 500 200]{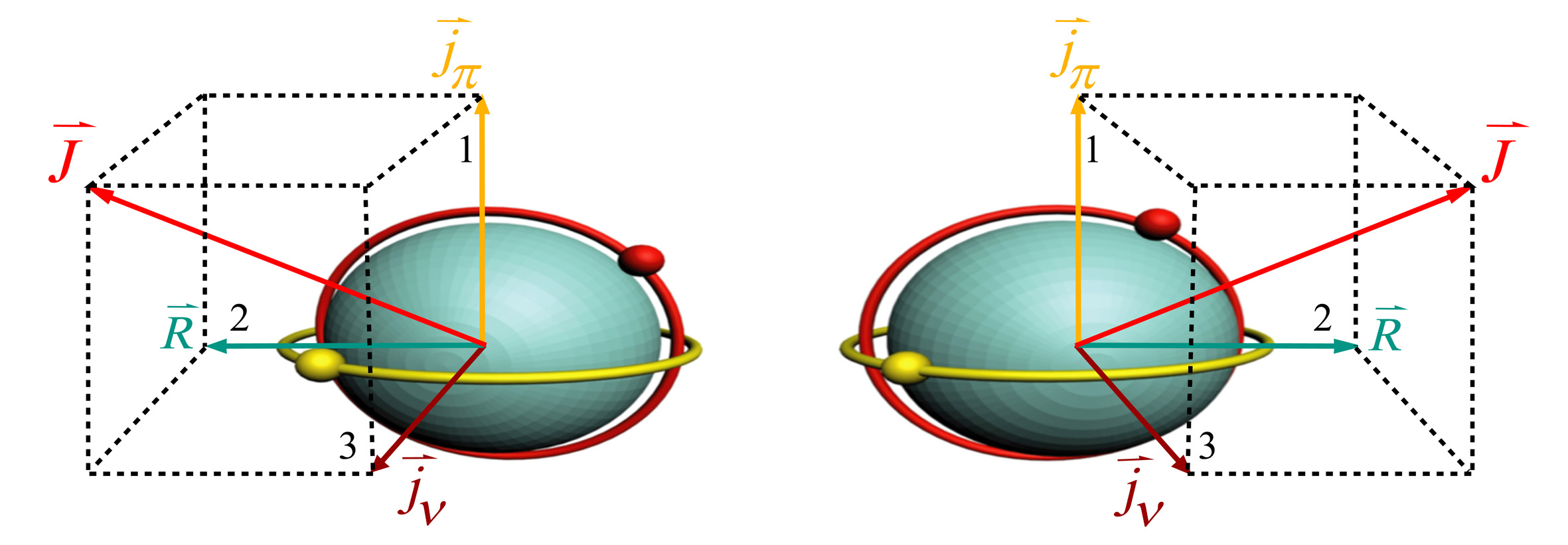}
\end{center}
\caption{Left- and right-handed chiral systems for a triaxial
odd-odd nucleus. }  \label{chiral}
\end{figure}

The spontaneous broken of chiral symmetry is expected to occur in an
atomic nucleus with a triaxial shape as well as a few high-$j$
valence particles and a few high-$j$ valence holes. For a triaxially
deformed rotational nucleus, the collective angular momentum favors
alignment along the intermediate axis, which in this case has the
largest moment of inertia, while the angular momentum vectors of the
valence particles (holes) favor alignment along the nuclear short
(long) axis. The three mutually perpendicular angular momenta can be
arranged to form two systems with opposite chirality, namely left-
and right-handedness, see Fig.~\ref{chiral} for a schematic
illustration. These two systems are transformed into each other by
the chiral operator which combines time reversal and spatial
rotation of 180$^\circ$, $\chi={\cal TR}(\pi)$. The spontaneous
breaking of chiral symmetry thus happens in the body-fixed reference
frame. In the laboratory reference frame, with the restoration of
chiral symmetry due to quantum tunneling, the so-called chiral
doublet bands, i.e., a pair of $\Delta I = 1$ bands (normally
regarded as near degenerate) with the same parity, are expected to
be observed in triaxial nuclei.

\subsection{Classic picture}

This chiral picture is firstly proposed and illustrated
theoretically in PRM and TAC approach~\cite{FM97}.

TAC is the version of the mean field theory that permits the
calculation of the orientation of the deformed density distribution
relative to the (space-fixed) angular momentum vector. For the
triaxial nuclei, as illustrated in Fig.~\ref{chiral}, there exists
the possibility of the aplanar solutions, where the angular momentum
vector does not lie in one of the principle planes defined by the
principal axes of the deformed density distribution. Due to the
invariance of the intrinsic deformed density distribution with the
rotations ${\cal R}_1(\pi), {\cal R}_2(\pi), {\cal R}_3(\pi)$($D_2$
symmetry), the aplanar TAC solutions are restricted in two of eight
octants of principle axes frame, i.e., the first octant
(right-handed) and the fourth octant (left-handed) shown in
Fig.~\ref{chiral}. These two aplanar solutions are chiral as they
cannot be transformed into each other by a rotation, but degenerate
as TAC is a semi-classic model which cannot describe the quantum
tunneling between the left-handed and right-handed
states~\cite{FM97}.

The mean field version of the TAC adopted in Ref.~\cite{FM97} is the
simple single-$j$ model. With a hybrid potential combining the
spherical Woods-Saxon single-particle energies and the deformed part
of the Nilsson potential, chiral rotation has been studied by the
Strutinsky shell correction TAC (SCTAC)
method~\cite{Dimitrov00PRL,Dimitrov00PRC}. More microscopically, the
self-consistent Skyrme Hartree-Fock cranking model has been
developed and provides chiral solutions in $N=75$
isotones~\cite{Olbratowski04,Olbratowski06}. The cranked
relativistic mean field (RMF) theory has been reported only in the
contexts of principle axis rotation~\cite{Koepf89,Afanasjev00} and
planar rotation~\cite{Madokoro00,PengJ08TAC}. The generalization
thereof for searching for chiral solutions, i.e., the aplanar
rotation, is still under development.

The advantage of the cranked mean field approach is that it can be
easily extended to the multi-quasiparticle case. However, the usual
cranking approach is a semiclassical model, where the total angular
momentum is not a good quantum number, and the description of
quantum tunneling of chiral partners is beyond the mean field
approximation~\cite{Frauendorf01}.

\subsection{Quantum picture}

PRM is a quantum model consisting of the collective rotation and the
intrinsic single-particle motions, in which the total Hamiltonian
are diagonalized with total angular momentum as a good quantum
number. The PRM describes a system in the laboratory reference frame
in which the spontaneous broken chiral symmetry in intrinsic
reference frame has been restored. The energy splitting and quantum
tunneling between doublet bands can be obtained directly. By
analyzing the orientations of the angular momenta for the rotor as
well as the valence proton and neutron, and the effective angles
between these angular momenta, the chiral geometry represented by a
remarkable and similar aplanar rotation between doublet bands can be
revealed in a quantum way~\cite{ZhangSQ07,WangSY08,Lawrie09}.

Chirality for nuclei in $A\sim100$ and $A\sim130$ regions has been
studied with the particle-rotor model~\cite{PengJ03,PengJ03cpl}, or
the core-quasiparticle/core-particle-hole coupling
model~\cite{Starosta02,Koike03} by following the
Kerman-Klein-D\"{o}nau-Frauendorf method~\cite{Klein00}. With the
pairing correlations taken into account to simulate the
configurations of multi-particles sitting in a high $j$-shell, PRM
with a quasi-proton and a quasi-neutron coupled with a triaxial
rotor has been applied to study chiral doublet
bands~\cite{ZhangSQ07,WangSY07}. The (quasi-)particle rotor model
can well describe all the energy spectra and the ratios
$B(M1)/B(E2)$ and $B(M1)_{\textrm{in}}/B(M1)_{\textrm{out}}$ of
chiral bands for most nuclei, for example $^{126}$Cs in
$A\sim130$~\cite{WangSY07}, $^{106}$Rh in
$A\sim100$~\cite{WangSY08}, and $^{198}$Tl in $A\sim190$ mass
region~\cite{Lawrie09}. Recently, a triaxial $n$-particle-$n$-hole
PRM to treat more than one valence proton and one valence neutron
has been developed and applied to the study of nuclear chirality in
$^{135}$Nd with the 2$p$1$n$ configuration $\pi
h^{2}_{11/2}\otimes\nu h^{-1}_{11/2}$~\cite{Qi09PLB}.

Although the particle rotor model has been applied to investigate
the nuclear chirality extensively, it is a phenomenological model
based on a rigid triaxially deformed core. The quadrupole
deformation parameters $\beta$ and $\gamma$ as well as the
particle-hole configuration are imposed from the beginning.
Therefore, the response of the deformation and triaxiality with the
nuclear rotation as well as their influence on chirality are beyond
the current version of PRM.

\section{Rigidity of Finite Many-Body System}

\subsection{Stable $\gamma$ deformation }

By means of PRM with stable $\gamma$ deformation, it has been
demonstrated that for the symmetric particle-hole configuration such
as $\pi h_{11/2}\otimes\nu h_{11/2}^{-1}$ in $A\sim130$ region and
$\pi h_{9/2}\otimes\nu h_{9/2}^{-1}$ in $A\sim80$ region, the best
condition for the appearance of the chiral doublet bands is the
maximum triaxiality $\gamma=30^\circ$, while for the asymmetric
particle-hole configuration such as $\pi g_{9/2}^{-1}\otimes\nu
h_{11/2}$ in $A\sim$100 region, the best condition deviates the
maximum triaxiality to $\gamma
\sim27^\circ$~\cite{PengJ03,PengJ03cpl}. Generally speaking, the
chiral doublet bands are expected in the interval
$20^\circ\leq\gamma\leq 40^\circ$ for PRM with particle-hole
configuration .

For PRM with quasi-particle configurations, further investigation
has verified that the chiral geometry holds for the same deformation
interval $20^\circ\leq\gamma\leq 40^\circ$~\cite{ZhangSQ07,Qi09PLB}.
However, the deviation of the shape from $\gamma=30^{\circ}$ or the
configuration from pure particle-hole case will retrogress the level
degeneracy and prefer a near constant energy difference between
doublet bands.

Naturally, one must go beyond the limit of rigid $\gamma$
deformation in the future study of the nuclear chirality. The
potential energy surface in $(\beta, \gamma)$ plane in the
microscopic mean field calculations including the triaxial degree of
freedom will provide hints on these issues.

\subsection{Shape coexistence and M$\chi$D}

The adiabatic and configuration-fixed constrained triaxial RMF
approaches were developed and applied to $^{106}$Rh~\cite{MengJ06}.
The triaxial shape coexistence as well as the $\beta$ and $\gamma$
deformation suitable for chirality were found. Based on the triaxial
deformations and their corresponding high-$j$ proton-hole and
neutron-particle configurations, the possible existence of more than
one pair of chiral bands in one single nucleus --- multiple chiral
bands (M$\chi$D) --- is suggested in $^{106}$Rh. Such investigation
has been extended to the rhodium isotopes and the existence of
M$\chi$D is suggested in $^{104,106,108,110}$Rh~\cite{PengJ08}. The
investigation provides not only further support for the prediction
of M$\chi$D in $^{106}$Rh, but also presents new experimental
opportunity for the observation of M$\chi$D in $A\sim100$ mass
region.

The prediction of M$\chi$D~\cite{MengJ06,PengJ08} has been further
examined by including time-odd fields~\cite{YaoJM09}. The
configuration-fixed constrained triaxial relativistic mean-field
approach including time-odd fields has been applied to study the
candidate M$\chi$D nucleus $^{106}$Rh. The calculations support the
previous prediction of the M$\chi$D, although the time-odd fields
contribute 0.1-0.3MeV to the total energy and slightly modify the
$\beta$ and $\gamma$ deformation.

So far in the constraint RMF calculations, the rotational degree of
freedom is switched off. In order to include the rotational degree
of freedom in microscopic energy density functional theories
self-consistently, the three-dimensional cranking or even angular
momentum projection techniques should be applied and the deformation
should be obtained by minimizing the energy for given total angular
momentum, which is still quite challenging even for modern advanced
computational facilities.

\subsection{Deformation response to rotation}

Total Routhian surface (TRS) calculations are often used to track
the deformation variation with rotation. The idea of the TRS is to
search for the energy minimum in the deformation space for a given
rotational frequency. Based on the Nilsson or the Woods-Saxon
phenomenological potentials and the Strustinsky shell correction
method, TRS calculations are mostly restricted to the principle axis
cranking only. Its extension for tilted axis cranking is simple in
idea but quite difficult in practice, i.e., for a given cranking
rotation frequency $\hbar\omega$ with the tilted angles $(\theta,
\varphi)$, the deformations $(\beta, \gamma)$ are firstly obtained
by minimizing the TRS, then this procedure should be repeated for
all tilted angles $(\theta, \varphi)$ and finally the TRS should be
minimized with $(\theta, \varphi)$. In such a way, the deformation
and the tilted angles as a function of the rotation frequency can be
obtained.

In comparison, the corresponding microscopic calculations based on
modern energy density functionals have the advantage that the
deformations $(\beta, \gamma)$ are self-consistently obtained for a
given cranking rotation frequency $\hbar\omega$ with the tilted
angles $(\theta, \varphi)$ and one needs only minimizing the TRS
with $(\theta, \varphi)$. However, the self-consistent microscopic
iteration can be quite time consuming. So far such calculations have
been realized for conventional energy density
functionals~\cite{Olbratowski04,Olbratowski06} and covariant energy
density functionals in two dimensions~\cite{Madokoro00,PengJ08TAC}.

\subsection{Coupling between vibration and rotation}

For finite many-body systems such as the atomic nucleus, the
deformation may never be rigid but more or less soft. Naturally in
PRM, one may consider the vibration degree of freedom of the core in
addition to the collective rotation, i.e., the collective
vibration-rotation model~\cite{Bohr75}.

Taking such a collective vibration-rotation model, the calculated
results for the rigid or the $\gamma$-unstable core are
qualitatively very similar for an odd-odd nucleus~\cite{Droste09}.
Similarly, describing the core by the Interacting Boson Model (IBM)
with O(6) dynamical symmetry~\cite{Tonev07}, the chiral bands have
been investigated and the dynamic chirality with shape fluctuation
is proposed.

\section{Fingerprints for Nuclear Chirality}

\subsection{Energy spectra}

Since the prediction of the chirality in atomic nuclei~\cite{FM97},
the appearance of a pair of near degenerate $\Delta I = 1$ bands
with the same parity has been taken as its fingerprint. Of course,
the definition of the ``near degenerate" is subjective. Furthermore,
it should be emphasized that the experimental signals suggested as
the fingerprints of the nuclear chirality are essentially based on
one particle and one hole coupled to a rigid triaxial rotor with
$\gamma=30^\circ$~\cite{FM97,Koike04,WangSY07b}. The ``near
degenerate" depends on the deformation, valent nucleon
configuration, and their couplings. Normally, the observed ``near
degenerate" energy is around 200 keV.

From the energy spectra, one can also extract other physical
observables such as spin-alignment and the energy staggering
parameter $S(I)=[E(I)-E(I-1)]/2I$ and use them as possible
fingerprints. So far, most of the proposed chiral doublet candidates
are mainly based on the observed near degenerate $\Delta I = 1$
doublets bands.

A systematic explanation by excluding the chirality will be very
difficult, if not impossible, for the observed chiral doublet
candidates in more than 30 nuclei.

\subsection{Electromagnetic transitions}

With more and more chiral doublet candidates proposed, it is quite
natural to have observable fingerprints other than the energy
spectra. The electromagnetic transition in doublet bands become a
hot topic in identifying the chiral bands.

Based on the $\pi h_{11/2}\otimes \nu h^{-1}_{11/2}$ configuration
coupled to a rigid triaxial rotor with $\gamma=30^\circ$, the
selection rule for electromagnetic transitions in the chiral bands
has been proposed~\cite{Koike04}, including the odd-even staggering
of intraband $B(M1)/B(E2)$ ratios and interband $B(M1)$ values, as
well as the vanishing of the interband $B(E2)$ transitions at high
spin region.

With the improvement of experimental techniques, lifetime
measurements for the doublet bands have been done and the $B(M1)$
and $B(E2)$ transition probabilities are
extracted~\cite{Tonev06,Grodner06,Mukhopadhyay07,Suzuki08,WangLL09}
and examined against the fingerprints of the chiral doublet bands.

Similar as the case of the energy spectra, the fingerprints of the
electromagnetic transition depend on the deformation, valent nucleon
configuration, and their couplings too. The dependence of these
fingerprints on  triaxiality have been investigated in PRM with $\pi
h_{11/2}\otimes \nu h^{-1}_{11/2}$ configuration~\cite{Qi09PRC}. It
is found that the $B(M1)$ staggering is associated strongly with the
characters of nuclear chirality, i.e., the staggering is weak in
chiral vibration region and strong in the static chirality region.
This result also qualitatively agrees with the recent lifetime
measurements for the doublet bands in $^{128}$Cs and $^{135}$Nd. The
former was claimed as a good example revealing chiral symmetry
breaking~\cite{Grodner06}, where the pronounced $B(M1)$ staggering
is exhibited. The latter was suggested to reveal the chiral
vibration motions for $I<41/2\hbar$ region~\cite{Mukhopadhyay07},
where the weak $B(M1)$ staggering is shown.

Nevertheless, a model independent selection rule for electromagnetic
transitions in the chiral bands is still missing.

\subsection{Other fingerprints}

Apart from energy spectra and electromagnetic transitions, the
spin-alignment and the energy staggering parameter are also used in
the discussion of the chiral pair bands. As the spin-alignment and
the energy staggering parameter are extracted from the energy
spectra, they inherit naturally the same uncertainty. Ideally, the
spin alignments, moment of inertia, and electromagnetic transition
probabilities are expected to be identical, or in practice very
similar for chiral pair bands~\cite{Petrache06}.

Examining against these criteria, the doublet bands of $^{128}$Cs
have been regarded as the best known example for the chiral symmetry
broken~\cite{Grodner06}. While the lifetime measurement performed
for doublet bands in $^{134}$Pr, which had been considered as the
best examples of chiral rotation due to their extremely small level
discrepancy between the doublet bands, stirred quite a lot of
debates on the interpretation of the chirality~\cite{Tonev06,
Petrache06}. Again, new and model independent fingerprints are
highly demanded.

\section{Interpretations Excluding Chiral Picture}

Although candidate chiral doublet bands have been proposed in a
number of odd-odd, odd-$A$ or even-even nuclei in the $A \sim 100,
130, 190$ mass regions, their identification is mainly based on the
observed energy spectra of doublet bands. Several competitive
mechanisms have been proposed to explain the pair of near degenerate
$\Delta I = 1$ bands with the same parity.

\subsection{$\gamma$ band}

In the collective vibration-rotation model, the so-called
$\gamma$-band, which has been observed in many deformed nuclei, has
the quantum numbers $K=2, n_\beta=0, n_\gamma=0$ and is built on the
quantum mechanical zero point vibration in the
$\gamma$-direction~\cite{Bohr75,Ring81}. For the doublet bands in
Cs, La, Pr and Pm $N = 75$ isotones~\cite{Starosta01}, the
interpretation of the yrare band as a $\gamma$ vibration coupled to
the yrast band has been ruled out because the $\gamma$-vibration
energies in this mass region are $\geq 0.60$ MeV, which can not
explain the observed small energy displacement ($<0.40$ MeV).
However, one can not rule out the interpretation of doublet bands in
a $\gamma$ vibration picture. Furthermore, the influence of the
$\gamma$ vibration on the chiral rotation should be treated more
carefully.

\subsection{Shape coexistence}

In order to explain the observed electromagnetic transitions for the
doublet bands in $^{134}$Pr~\cite{Tonev06}, the concept of shape
coexistence has been proposed~\cite{Petrache06}. Based on a
two-bands crossing model, a  ratio around two for the transition
quadrupole moments has been obtained for the doublet bands in
$^{134}$Pr in the spin interval $I =$ 14--18  where the observed
energies are almost degenerate, which implies a considerable
difference in the nuclear shape associated with the two bands and
thus the doublet bands in $^{134}$Pr cannot be interpreted as chiral
bands~\cite{Petrache06}.

To reproduce the observed near degenerate spectra and
electromagnetic transitions in $^{134}$Pr with either the shape
coexistence or the chiral picture in a microscopic and
self-consistent way remains to be a challenge.

\subsection{Many particle correlations}

In the framework of the projected shell model (PSM) with triaxial
deformation, the doublet bands in $^{134}$Pr are
investigated~\cite{Chen06}. By reproducing the energy spectra, it is
found that in the region $I=$14 to 18, the band 1 is a
2-quasi-particle state, e.g., of $\pi h_{11/2}\nu h_{11/2}$
configuration, while the band 2 is mainly a 4-quasi-particle state,
e.g., of $\pi h_{11/2}d_{5/2}g_{7/2}\nu h_{11/2}$ configuration.
Therefore many particle correlations was proposed as another
possible mechanism of doublet bands in $^{134}$Pr~\cite{Chen06}. It
will be quite interesting to have more detailed and quantitative
comparisons with the observed energy spectra, $B(E2)$ and $B(M1)$
values in order to figure out the role of the many particle
correlations. In fact, the many particle correlations are also
necessary to be taken into account in PRM study of the nuclear
chirality.

\subsection{Pseudospin partner bands}

The pseudospin symmetry~\cite{Arima69,Hecht69} in finite nuclei have
been known as a relativistic symmetry and become of the hot topic in
current nuclear physics
frontiers~\cite{Ginocchio97,Meng98,Meng99,ChenTS03}. The concept of
the pseudospin symmetry has been used in the interpretation of the
identical bands observed in superdeformed nucleus~\cite{Zeng91}.

Recently, pairs of $\Delta I=1$ doublet bands with the same parity
and near degenerate energies have been observed in several odd-odd
and odd-$A$ nuclei,e.g., $^{108}$Tc~\cite{Xu08},
$^{128}$Pr~\cite{Petrache02}, $^{186}$Ir~\cite{Cardona97}, and
$^{195}$Pt~\cite{Petkov07}, and explained in terms of the coupling
of a proton (neutron) and a neutron (proton) pseudospin doublet.
Normally, the observed pseudospin doublet bands start at relatively
lower spin and show the opposite odd-even phase in the $B(M1)$
staggering between the partners, while the chiral doublet bands hold
the same phase.

\subsection{Core polarization}

The classic examples for chirality are triaxial odd-odd nuclei where
the angular momenta of a high-$j$ particle and a high-$j$ hole are
aligned along the short and long axis, respectively, and the angular
momentum of collective rotation is aligned with the intermediate
axis. The observation of doublet bands in odd-$A$ and even-even
nuclei further exemplify the general geometric character of chiral
symmetry breaking, because the non-planar rotation is related to
more complicated particles and holes configuration.

Another interesting issue is the core polarization effect. In
addition to the PRM and TAC approaches, the interacting boson fermi
model~\cite{Tonev07,Brant08,Brant09} and the pair truncated shell
model~\cite{Higashiyama05} have been applied to investigate the
observed doublet bands as well. In comparison with the PRM with a
rigid rotor, the quadrupole collective excitations of the even-even
core are taking into account by the ingredients of the collective
nucleon pairs with angular momenta of zero and two.

\section{Microscopic Model Expected}

\subsection{Beyond TAC}

Chiral rotation has been extensively studied by TAC method with a
phenomenological Woods-Saxon or Nilsson potential, and the
self-consistent Skyrme Hartree-Fock model. The advantage of the TAC
is that it can be easily extended to the multi-quasiparticle case.
However, the usual cranking approach is a semiclassical model, where
the total angular momentum is not a good quantum number and the
electromagnetic transitions are calculated in a semiclassical
approximation. Furthermore, the description of nonlinear quantum
tunneling of chiral partners is beyond the mean field approximation,
as well as the random phase approximation in which the excitations
of the equilibrium mean field are described as harmonic vibrations
only.

Before the onset of chirality, the precursor of the symmetry
breaking occurs as a soft vibration between the right- and
left-handed configurations. These chiral vibrations have been
studied in the framework of the random phase approximation (RPA)
based on the TAC mean field~\cite{Almehed08,Mukhopadhyay07}.
Currently, the RPA calculations are only limited on the TAC with a
spherical Woods-Saxon potential for the mean field together with a
QQ-force and a constant pair gap. Further investigation in a more
microscopic way is expected. In particular, we need a model which
could simultaneously describe the chiral vibration dominated by a
slow vibration, static chirality by the quantum tunneling, and the
smooth transition from chiral vibration to static chirality.

\subsection{Rotational symmetry restoration}

Due to the mean field approximation, which is rooted in both shell
models and microscopic approaches, the rotational symmetry has been
broken in the deformed intrinsic reference frame. In order to
restore the rotational symmetry, it is necessary to use the standard
angular momentum projection (AMP) techniques~\cite{Ring81}.

In the late 1970s, Hara and Iwasaki applied AMP techniques and
developed the projected shell model (PSM) to obtain the good angular
momentum states from the Nilsson state~\cite{Hara79,Hara95}. Later,
the triaxiality has been taken into account and applied to study the
yrast and $\gamma$-vibrational bands for both $\gamma$-soft and
well-deformed nuclei~\cite{Sheikh99,Gao06, Sheikh08}. In principle,
the doublet bands in triaxial nuclei can be understood in the
framework of PSM with triaxially deformed multi-quasiparticle
states. Such calculation has been reported for $^{134}$Pr, where the
partner bands have been suggested to have completely different
configurations~\cite{Chen06}. It will be very interesting to apply
the PSM to study the chiral doublet candidates systemically and
examine their detailed observables including the electromagnetic
properties and the energy spectra.

Due to the numerical complexity, only recently it become possible to
apply the AMP procedures for the microscopic energy density
functional~\cite{Valor00,Guzman02a,Guzman02b,Niksic06a,Niksic06b}.
However, in these studies, the axial symmetry has been imposed from
the beginning. Such restriction simplifies the numerical calculation
considerably, because the integrals over two of the three Euler
angles can be treated analytically and one is left with a
one-dimensional numerical integration.

In the context of energy density functionals, a full
three-dimensional angular momentum projection (3DAMP) has been
performed with a simple Skyrme-type interaction~\cite{Baye84}, and
the full Skyrme energy functional~\cite{Zdunczuk07}. Only very
recently, angular-momentum and particle-number projections with
configuration mixing have been attempted in the context of the
triaxial Hartree- Fock-Bogoliubov (HFB) theory with the full Skyrme
energy functional~\cite{Bender08}. On top of the triaxial
relativistic mean-field calculations, a full three-dimensional
angular momentum projection has been implemented~\cite{Yao09a}, even
with configuration mixing~\cite{Yao09b}. Although these modern
recipes have been successfully applied to the light even-even nuclei
like $^{24}$Mg, there is still a long way for their applications to
chiral doublet candidates due to the configuration space and the
quasi-particle excitations.

\subsection{Collective Hamiltonian parameter}

In this subsection, we would like to discuss the possibility for the
PRM to include the vibration-rotation coupling in a microscopic way
to study the nuclear chirality.

Recently, a new implementation is developed for the solution of the
eigenvalue problem of a five-dimensional collective Hamiltonian for
quadrupole vibrational and rotational degrees of freedom, with
parameters determined by constrained self-consistent relativistic
mean-field calculations for triaxial shapes. The model is tested in
a series of illustrative calculations of potential energy surfaces
and the resulting collective excitation spectra and transition
probabilities of the chain of even-even gadolinium~\cite{Niksic09},
neodymium and samarium isotopes~\cite{Li09}. For neodymium and
samarium isotopes, the first-order nuclear quantum phase transition
has been demonstrated in the characteristic energy spectra thus
obtained~\cite{Li09,Li09b}.

The possible extension of the PRM is the replacement of the core by
the collective Hamiltonian with the parameters extracted in a
similar way, which may provide a new dimension in the description of
the chiral doublet bands.

\section{Summary}

The spontaneous broken of chiral symmetry suggested in atomic
nucleus one decade ago has attracted lots of attention both
experimentally and theoretically. Several fingerprints of the
chirality in atomic nuclei have been proposed although ambiguities
still exist. Semi-classically or quantum mechanically, the
phenomenological or microscopic model has achieved great success in
prediction and the description of the chiral doublet bands. However,
the interpretation of the experimentally observed doublet bands as
chiral partners can be contradictory.

Theoretically, open problems in existing model are discussed,
including the description of quantum tunneling in the mean field
approximation as well as the deformation, core polarization and
configuration of PRM. More efforts are needed in developing a model
which could simultaneously describe the chiral vibration dominated
by a slow vibration, static chirality by the quantum tunneling, and
the smooth transition from chiral vibration to static chirality.
With the rapid development of computing facilities, a full
three-dimensional angular momentum projection on top of the triaxial
relativistic mean-field calculations with quasi-particle excitation
may be realized. Meanwhile the PRM can also be combined with the
collective Hamiltonian with the parameters extracted in a
microscopic way.

\section*{Acknowledgments}
The authors wish to thank Y.S. Chen, Z.C. Gao, Z.P. Li, B. Qi, F.R.
Xu, J.M. Yao and S.G. Zhou for carefully reading the manuscript and
useful discussions. This work is partly supported by the Major State
Basic Research Developing Program 2007CB815000, the National Natural
Science Foundation of China under Grant Nos. 10775004, 10975007,
10975008 and 10720003.

\end{document}